# Link the Fluorescence and TEM Studies on Amyloid Fiber Formation -- To Reconstruct the Fiber Length Distribution Based on the Knowledge of Moments Evolution

Liu Hong*

Zhou Pei-Yuan Center for Applied Mathematics, Tsinghua University, Beijing, China

*zcamhl@tsinghua.edu.cn

**Abstract:** In the study of amyloid fiber formation, the fiber length distribution plays a central role. From this microscopic distribution, all macroscopic statistical quantities, like the fiber number concentration $P(t)$ and mass concentration $M(t)$, can be easily achieved. However the inverse problem is usually non-trivial. In this paper, a practical method is introduced to show how to reconstruct the fiber length distribution function based on the knowledge of $P(t)$ and $M(t)$. Compared to directly solving mass-action equations, our method speeds up the calculation by at least ten thousand times (from days to seconds). And the accuracy is also quite satisfactory. Furthermore our method provides an interesting linkage between the fluorescence imaging and more detailed TEM studies, which may inspire wider applications in this field.

**Keywords:** amyloid fiber, fiber length distribution, ThT fluorescence, TEM, inverse problem

## 1. Introduction

The formation of amyloid fiber, caused by protein misfolding or unfolding, is a typical self-assembling process of macromolecules[1]. Due to its significant correlation with many well-known neurodegenerative diseases[2,3], it has became a hot topic in molecular biology and neureoscience nowadays.

To quantitatively characterize the formation process of amyloid fiber, the fiber length distribution plays a central role[4,5], from which all experimentally measurable and theoretically interesting quantities can be derived. With the development of modern microscopic visualization technology, now we can directly observe the high-resolution image of fiber structure and count their frequency by TEM, STM and so on[6,7]. However the expensiveness and complexity of these equipments limit their usage in the study of amyloid fiber formation. On the other hand, the fluorescence imaging (especially the ThT fluorescence) has became the most popular method in this field[8,9]. Thioflavin T (ThT) is known as a reagent which can bind rapidly to amyloid fibrils (especially the beta-sheet rich parts) and become strongly fluorescent[9]. And the emitted fluorescence intensity is believed proportional to the mass concentration of fibrils (the first-order moment of fiber length distribution) in the system[10]. This dramatic fact makes ThT a useful probe for the real-time monitoring of the fiber formation.

If the fiber length distribution is recorded in real-time, the calculation of fiber mass concentration is trivial. Contrarily, if the time-evolution data of fiber mass concentration is known, to determine the fiber length distribution is not an easy task. This raises an interesting question: Can we predict the TEM data (or other high-resolution miroscopic visualization tools) about fiber length distribution based on the knowledge of ThT fluorescence intensity measurements both efficiently and accurately?

To solve this problem, here a systematic method is introduced. To be exact, we will show how to derive model parameters (mainly the reaction rate constants) from the evolutionary curves of ThT fluorescence intensity, how to reconstruct the fiber length distribution based on the knowledge of number and mass concentration of fibrils, and so on. Instead of seeking for the mathematical strictness, we try to achieve a balance between the accuracy and efficiency of the mehtod. It should be easy, fast and of enough accuracy to make quantitative comparisons with experimental data. So in some sense it is a "practical" method and can be directly applied to most real cases.

The validity of our method lies on following facts:

(1) The formation of amyloid fiber can be characterized through simple kinetic models in the form of mass-action equations, which include nucleation, elongation and fragmentation[11,12,13]. This is the working foundation of our method.

(2) Without directly solving the mass-action equations, which is generally high dimensional and thus time consuming, the time evolution of number concentration and mass concentration of fibrils can be predicted with high precision through simple moment-closure equations constituted by only two equations[14,15]. This is the key to the high efficiency of our method.

(3) The scaling relations for the kinetic and equilibrium quantities can largely reduce the possible parameter space and thus will greatly accelerate the determination procedure of model parameters[12,14]. This part also helps to raise the performance.

(4) The real fiber length distribution can be approximated through simple empirical formula[16,17,18], in which the unknown fitting parameters can be uniquely determined based on the knowledge of moments (*e.g.* number concentration and mass concentration of fibrils). As there is no general guidelines for the choice of the approximate fiber length distribution function, this step will be the trickiest part and largely rely on the experience of the researchers.

## 2. Model and Method

The basic procedure to interpret the TEM data about fiber length distribution based on the knowledge of ThT fluorescence intensity can be separated into four key steps: (1) extract kinetic and equilibrium characteristic quantities from the measurements of ThT fluorescence intensity; (2) build a proper model and link the characteristic quantities with model parameters in quantity; (3) Simulate moment-closure equations and find suitable model parameters through fitting fluorescenece data; (4) construct an approximate fiber length distribution with the knowledge of moments and interpret the TEM data. The details of each step are given as follows.

### (1) Extract characteristic quantities

The measured intensity curves for ThT fluorescence, which are evolved in time, can generally be fitted by following empirical formula, *i.e.*[19]

$$F(t) = F(0) + \frac{F(\infty) - F(0)}{1 + \exp\left[-4k_{1/2}\left(t - t_{1/2}\right)\right]}, \qquad (1)$$

where $F(t)$ is the ThT fluorescence intensity measured at time $t$; $F(0)$ and $F(\infty)$ are the measured intensity at the initial and final time.

In above formula, two most important quantities, which determine the major kinetic behaviors of amyloid fiber formation, are the half-time of fiber growth $t_{1/2}$ (defined as the time point when fiber mass concentration reaches half of the total value $M(t_{1/2}) = m_{tot}/2$) and the apparent fiber growth rate $k_{1/2}$ (defined as the normalized fiber growth rate at the half time $k_{1/2} = \dot{M}(t_{1/2})/m_{tot}$). Note here an extra factor "4" is added to Eq. 1 to keep the definition of apparent fiber growth rate consistent.

The information of extracted characteristic quantities can help us to pick a correct model and determine some model parameters (especially the critical nucleus size and the scaling exponent) in the next step.

**(2) Build model and derive constraint relations for model parameters**

As many previous studies have shown, the formation process of amyloid fiber can be characterized by simple kinetic models, which are formulated in a form of mass-action equations[11,12,13]. Without loss of generality, here we consider a model includes three basic processes: primary nucleation, elongation and fragmentation (and their correponding backward reactions). It has been approved very successful in modeling breakable filaments and also contains non-breakable fibrils as a special case (when $k_f^+ = 0$)[12,13,14].

$$n_c A_1 \underset{k_n^-}{\overset{k_n^+}{\rightleftharpoons}} A_{n_c},$$

$$A_1 + A_i \underset{k_e^-}{\overset{k_e^+}{\rightleftharpoons}} A_{i+1}, \qquad (i \geq n_c) \qquad (2)$$

$$A_{i+j} \underset{k_f^-(i,j)}{\overset{k_f^+(i,j)}{\rightleftharpoons}} A_i + A_j, \qquad (i, j \geq n_c)$$

where $n_c$ stands for the size of critical nucleus. The reaction rates for fiber fragmentation and association are chosen according to Hill's model as[20]

$$\begin{cases} k_f^+(i,j) = k_f^+ (ij)^{n-1} (i \ln j + j \ln i)/(i+j)^{n+1}, \\ k_f^-(i,j) = k_f^- (i \ln j + j \ln i)/[ij(i+j)], \end{cases} \qquad (3)$$

where the scaling exponent $n$ represents degrees of freedom of a fiber in water solution.

To solve above model by mass-action equations[21], determining all unknown reaction rate constants is of first importance. Using the kinetic characteristic quantities derived in the first step, we can correlate the model parameters by following quantitative relations(reported in Ref. [15]), which will largely reduce the possible parameter space and thus accelerate the fitting procedure.

$$\begin{cases} k_{1/2} \sim \left[ \left( k_e^+ m_{tot} \right)^{n-1} k_f^+ \right]^{1/n}, \\ t_{1/2} \cdot k_{1/2} \sim \ln \left[ \left( k_f^+ \right)^{2/n} \Big/ \left( k_n^+ m_{tot}^{n_c-1} \right) \right]. \end{cases} \qquad (4)$$

In most cases, we can choose the critical nucleus size $n_c = 2$ (some may require a larger value, e.g. $n_c = 5$ for Apo C-II in Ref. [12]) and the scaling exponent for fragmentation $n = 0 \sim 4$. Their exact values can be derived from the scaling dependence of $t_{1/2}$ and $k_{1/2}$ on the total protein concentration $m_{tot}$.

**(3) Simulate moment-closure equations**

To interpret the TEM data in real time, the time evolution of fiber length distribution should be known, which in principle can be directly calculated by mass-action equations[4,5]. However in most cases this is an unpleasant way, since the dimension of the ODE (ordinary differential equations) system involved is very large ($10^3 \sim 10^6$) and the calculation will be extermely time-consuming. To conquer this difficulty, here an alternative approach is adopted, that is: firstly calculate the time evolution of number and mass concentration of filaments; then reconstruct the fiber length distribution based on the knowledge of $P(t)$ and $M(t)$ [22].

Thanks to the moment-closure method, the fisrt step has been achieved in our previous studies[15]. Without solving original mass-action equations, we can directly determine $P(t)$ and $M(t)$ through following moment-closure equations,

$$\begin{cases} \dfrac{dP}{dt} = k_n^+ \left(m_{tot} - M\right)^{n_c} - k_n^- \left(1-\theta\right) P + k_f^+ \left(1-\theta\right) P \theta^{n_c} \Xi_{n-1} - k_f^- \left(1-\theta\right)^2 P^2 \Lambda_1, \\ \dfrac{dM}{dt} = n_c k_n^+ \left(m_{tot} - M\right)^{n_c} + 2k_e^+ \left(m_{tot} - M\right) P - 2k_e^- P - \left(n_c k_n^- - 2k_e^-\right)\left(1-\theta\right) P, \end{cases} \quad (5)$$

where $\theta \equiv (M - n_c P)/[M - (n_c - 1)P] \in [0,1)$. And the empirical functions are

$$\Xi_{n-1} = \dfrac{n_c^n}{2^n} \left[ \dfrac{\ln n_c}{n_c^2(-\ln\theta)^2} + \dfrac{(n-2)\ln n_c + 1}{n_c^3(-\ln\theta)^3} + \dfrac{(3n-5)(n-4)\ln n_c + 6(n-3)}{4n_c^4(-\ln\theta)^4} + \cdots \right] \quad \text{and}$$

$$\Lambda_1 = \dfrac{\ln n_c}{n_c^2(-\ln\theta)^2}.$$

Bearing the relations for model parameters (Eq. 4) in mind, then after not too many times of tries, we can find a group of parameters which fit the intensity data of ThT fluorescence satisfactorily. Comparing to direct calculation of mass-action equations (at least in days), the current method has accelerated the fitting procedure by at least ten thousand times (in seconds) and thus makes the data analysis in real-time be possible.

**(4) Construct the fiber length distribution**

The final and the most crucial step is to reconstruct the fiber length distribution based on the knowledge of $P(t)$ and $M(t)$, which have been calculated through momemt-closure equations in the last step. Instead of seeking for an explicit form for the fiber length distribution function, we try to achieve a balance between the precision and the simplicity. Inspired by the solution obtained under the condition of detailed balance between fiber fragmentation and association ($k_f^+(i,j)[A_{i+j}] = k_f^-(i,j)[A_i][A_j]$), we assume the approximate function for fiber length distribution satisfies following form, i.e.

$$[A_i](t) = i^n [A_1]^i (t) \exp\left(-\varepsilon_f + i\varepsilon_m\right), \qquad (i \geq n_c) \quad (6)$$

where the fitting parameters $\varepsilon_f$, $\varepsilon_m$ and $[A_1]$ can be determined through following normalization conditions

$$\begin{cases} \sum_{i=n_c}^{\infty} i^n [A_1]^i \exp\left(-\varepsilon_f + i\varepsilon_m\right) = P, \\ \sum_{i=n_c}^{\infty} i^{n+1} [A_1]^i \exp\left(-\varepsilon_f + i\varepsilon_m\right) = M, \\ [A_1] + M = m_{tot}. \end{cases} \quad (7)$$

When $n > 0$, the fiber length distribution in Eq. 6 shows a bell shape with the maximum located at $i = n/(\varepsilon_m - \ln[A_1])$, which turns to be a reasonable candidate for breakable amyloid fibrils[23,24]. But if $n = 0$ (the fiber immobile case), Eq. 6 turns into an exponentially decaying function, exactly the same as the one studied by Dill *et al.*[25] and suitable for non-breakable fibrils. Other notable empirical formulas for fiber length distribution reported in the literature include: the form $[A_i] = ai^{b-1}\exp(-ai^b)$ proposed by Fu *et al.*[16], where $a$ and $b$ are size and shape parameters respectively; and the log-normal distribution function $[A_i] = \exp\left[-(\ln i - \mu)^2/(2\sigma^2)\right]/(i\sigma\sqrt{2\pi})$, where $\mu$ and $\sigma$ are the mean and standard derivation of the logarithm of fiber length[17]. These forms may be useful in other applications and will not be discussed here.

### 3. Applications

As a concrete example, here we apply our method to the experiments of MoPrP fibrils performed by Luo *et al.*[26]. From the measured evolutionary curve for ThT fluorescence intensity (Fig. 2A), we can determine the half-time for fiber growth as $t_{1/2} = 9.36 \times 10^4 s$ and the apparent fiber growth rate as $k_{1/2} = 8.89 \times 10^{-6} s^{-1}$. Furthermore form the TEM imaging, the average fiber length is estimated as ~912 monomers[27]. For simplicity, we firstly set $k_n^- = k_e^- = 0$. Then using the relations in Eq. 4[28], we simulate Eq. 5 and easily determine a group of model parameters which provides a best fitting of the ThT fluorescence intensity curve (Fig. 2A). Consequently with the calculated values for $P(t)$ and $M(t)$, we reconstruct the fiber length distribution and the cumulative frequency based on Eq. 6 and 7, which turn to match the data measured by TEM very well (Fig. 2C and 2D).

The validity of our method is further confirmed by a direct comparison of the approximate fiber length distribution (Eq. 6) and the exact one calculated through mass-action equations in real time (Fig. 2E). We can see that except for early time perriods (within 16 hours), in which the fiber growth is still dominated by primary nucleation and elongation, and fragmentation can be nelgected due to the too low concentration of fibrils, the fiber length distribution function proposed in Eq. 6 turns to be a very good approximation to the exact one. Thus for most usual applications, we can safely adopt above method and expect satisfactory results in the end. However for very elegant studies, further efforts have to be made on the construction of better empirical formulas for fiber length distribution.

### 4. Discussions

In this paper, we introduce a practical method to show how to predict the high-resolution TEM (or STM) data about fiber length distribution based on the knowledge of limited information on fiber mass concentration obtained from measurements of ThT fluorescence intensity step by step. The major advantage of our method lies on its astonishing efficiency. Comparing to direct calculation of mass-action equations, our method raises the performance by as least ten thousand times (from days to seconds). If further counting the tedious data fitting procedure, our results become much more encouraging. Besides this method also provides an interesting link between the measurements of fluorescence intensity and more detailed TEM or STM studies, which may inspire wider applications in this field.

The necessity of the first and the second step is often easily ignored. However, in real applications, they are very useful and serve as a preliminary to following steps. The information of extracted characteristic quantities can not only help us to make a choice of which model is suitable for current data set, but also facilitate the determination of the critical nucleus size and the scaling exponent for fragmentation. While the quantitative relations in Eq. 4 enjoys great importance in reducing the possible parameter space and accelating the data fitting procedure. In a pity, currently we have only used the scaling form (the weak form). To make them as equalities (the strong form), further efforts must be made to determine all unknown coefficients. There is no complete answer yet.

Another notable point is related to the fourth step. As we have mentioned, to decide what kinds of approximate fiber length distribution function to use is very tricky and largely relies on the experience. Our current choice is certainly not the best. Other formulas could be tried, for example the log-normal distribution. In that case, corresponding changes should be made on the moment-closure equations and the normalization conditions, but the basic idea of our method will not be affected at all.

**Acknowledgements**

This work was partially supported by the National Natural Science Foundation of China(NSFC 11204150) and by Tsinghua University Initiative Scientific Research Program (20121087902).

generally give wrong values of other quantities) are found, we can vary them according to above relations to finetume those unsatisfied quantities while keeping $t_{1/2}$ and $k_{1/2}$ unchanged. This trick can largely reduce the workload in data fitting.

## Figure Legends

Figure 1. An illustration of the central idea of our method.

Figure 2. Comparison of the experimental data and numerical calculations on MoPrP fibrils. (A, B) Time evolution curves for the mass concentration and number concentration of MoPrP fribrils. The measured data of ThT fluorescence intensity in Ref. [23] are notated by black dots; while the numerical solutions of mass-action equations and moment-closure equations (Eq. 5) are given through red solid lines and blue dashed lines respectively. (C) Comparison of the solutions of mass-action equations (red solid lines) and our method (Eq. 5+Eq. 6) (blue dashed lines) on the fiber length distribution in real-time. (D, E) Comparison of TEM data in Ref. [23] (red bars) and the data obtained by our method (blue dashed lines) on the fiber length distribution and the cumulative frequency at 72h. The model parameters are set as $m_{tot} = 4 \times 10^{-5} M$, $n_c = 2$, $n = 4$, $k_n^+ = 9.1 \times 10^{-5} M^{-1} \cdot s^{-1}$, $k_e^+ = 550 M^{-1} \cdot s^{-1}$, $k_f^+ = 6 \times 10^{-14} s^{-1}$, $k_f^- = 55 M^{-1} \cdot s^{-1}$ and $k_n^- = k_e^- = 0$.

## Figures

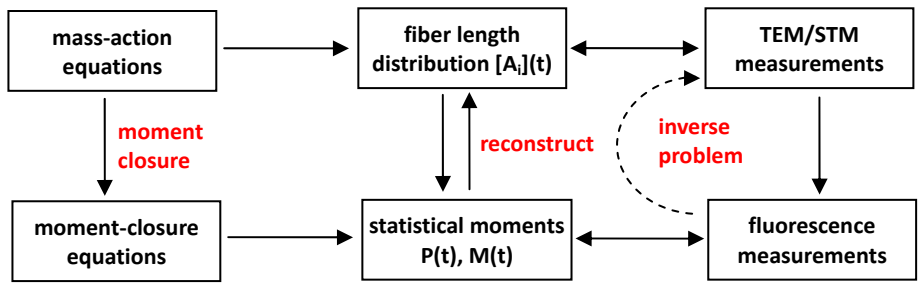

Figure 1

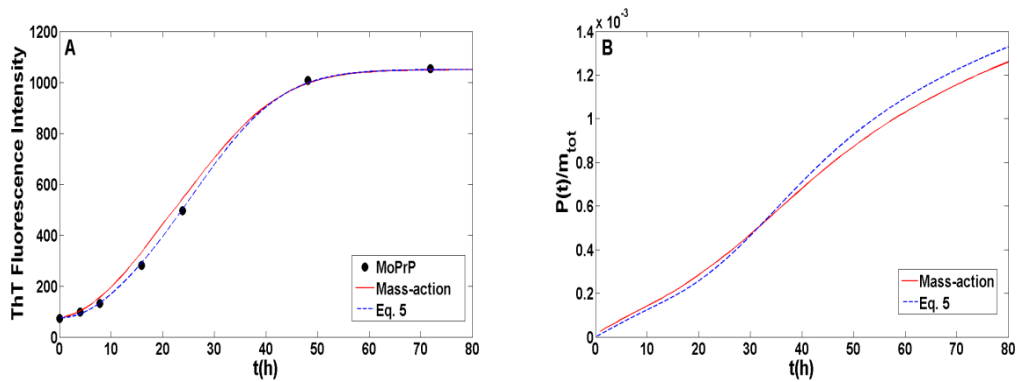

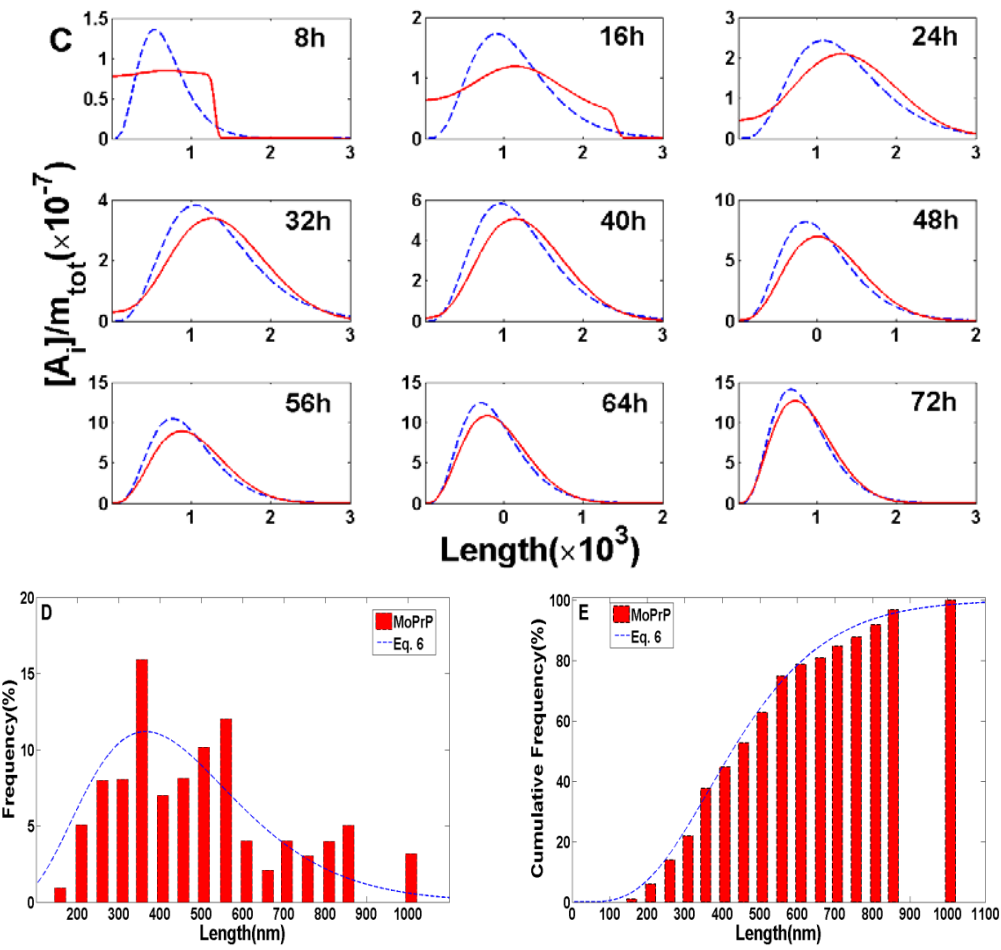

Figure 2